\documentclass[aps,prx,twocolumn,reprint,superscriptaddress]{revtex4-1}
\usepackage{graphicx}
\usepackage{physics}
\usepackage{float}
\usepackage{caption}
\usepackage{hyperref}
\usepackage{mathtools}
\usepackage{breqn}
\usepackage{amsfonts}
\usepackage{amsmath}
\usepackage{mathtools}
\usepackage{xcolor}
\begingroup\makeatletter
\catcode`,=\active
\global\let\breqn@comma,
\protected\gdef,{\ifmmode\expandafter\breqn@comma\else\expandafter\active@comma\fi}
\endgroup

\newcommand{\vect}[1]{\boldsymbol{#1}} 
\usepackage[nameinlink,capitalize]{cleveref}
\def\Arg{\mathop{\operator@font Arg}\nolimits}
\usepackage{amsmath}

\begin{document}
\title{Quantum option pricing using Wick rotated imaginary time evolution}
\author{\href{https://www.santoshkumarradha.me}{Santosh Kumar Radha}}
\email{Corresponding author:santosh@agnostiqlabs.com}
\affiliation{Agnostiq Inc., 180 Dundas Street W, Suite 2500, Toronto, ON M5G 1Z8, Canada}
\affiliation{Department of Physics, Case Western Reserve University, 10900 Euclid Avenue, Cleveland, OH-44106-7079, USA}


\begin{abstract}
In this paper we reformulate the problem of pricing options in a quantum setting.  Our proposed algorithm involves preparing an initial state, representing the option price, and then evolving it using existing imaginary time simulation algorithms. This way of pricing options boils down to mapping an initial option price to a quantum state and then simulating the time dependence in Wick's imaginary time space. We numerically verify our algorithm for European options using a particular imaginary time evolution algorithm as proof of concept and show how it can be extended to path dependent options like Asian options.  As the proposed method uses a hybrid variational algorithm, it is bound to be relevant for near-term quantum computers.
\end{abstract}
\maketitle
\section{Introduction}

Numerical methods are commonly used for pricing financial derivatives and are also used extensively in modern risk management. In general, advanced financial asset models are able to capture nontrivial features that are observed in financial markets. Due to the probabilistic nature of these assets, the calculation of their fair price based on existing information from the markets is a valuable business problem.  The seminal work of Black and Scholes ushered in a new era of the option pricing theory \cite{black1973pricing} after which the notion of option pricing models expanded rapidly and attracted considerable attention. More often than not, these asset pricing models are  multi-dimensional and complex, as a consequence of which, one does not have closed-form solutions. 
As is the case for any complex problem that  lacks a closed-form solution, extensive effort has been devoted to development of  alternative  solutions to the option pricing problem. 

A wide range of studies have focused on the numerical realization of the option pricing problem, ranging from stochastic simulations to numerical solutions of partial differential equations. Numerical techniques for pricing options can be classified into three main categories: Tree methods \cite{cox1979option,rendleman1979two}, Partial Difference methods\cite{reisinger2018finite} and Monte-Carlo methods\cite{boyle1977options,mc1,mc2}. The Binomial Tree method is simple  to implement and provides reasonably accurate results.  However, it is mainly used for pricing ``simple" options, with constant volatility.
The major drawback of the binomial method is that it is computationally resource intensive; it requires even more than what is needed for other methods that give comparably accurate results\cite{broadie1996american}. Because of this, it is not applied to more complex derivatives. Since asset pricing is inherently probabilistic and is described by stochastic differential equations,  Partial Difference methods can be used to price  financial derivatives by solving the stochastic differential equations (SDE) that the latter satisfies.  Finally in Monte-Carlo methods, one produces a large number of paths that the price of the underlying asset could follow in subsequent time steps using the solution of the SDE that characterizes it. When these simulations were initially presented as a forward-looking technique, it had been impossible to value American options and ,in general, path-dependent options. A large body of work has since been devoted to developing new approaches that enable Monte Carlo simulations to implement backward-looking algorithms which solve this inadequacy\cite{grant1997path,ibanez2004monte}. In order to determine the optimal exercise strategy, some additional numerical procedures must be embedded in the Monte Carlo method when pricing early exercisable products, e.g. American options, Bermudan options, \textit{etc}. Monte-Carlo methods usually lack efficiency, while the other two methods are plagued by the curse of dimensionality in high-dimensional settings.  Therefore, industries like finance are in constant search for novel  computing paradigms that might serve solutions to these computationally expensive but industrially valuable problems. There has also recently been a lot of interest in leveraging the power of machine learning tools, mainly the use of neural networks, to solve various forms of option pricing\cite{de2018machine,ferguson2018deeply,hahn2013option,mcghee2018artificial}.

Quantum computing promises to solve many of the existing problems that are otherwise infeasible to solve using classical computers. Today's quantum computers belong to the Noisy Intermediate-Scale Quantum (NISQ)\cite{Preskill2018} regime and are plagued with various shortcomings such as decoherence and   read-out errors that limit the applicability of this technology to practical problems. Although not ideal, algorithms exploiting the power of NISQ era quantum computers have provided new ways of solving computationally intensive problems in fields such as  machine learning\cite{Biamonte2017,Havlek2019,gao2018quantum} and quantum chemistry\cite{PhysRevApplied.11.044092,Kandala2017}. NISQ friendly quantum algorithms have also been proposed in hybrid settings as subroutines for computationally resource intensive parts of  classical algorithms\cite{moll2018quantum,parrish2019jacobi}. 

Quantum algorithms for solving computationally hard financial problems have recently been proposed for topics ranging from portfolio optimization\cite{rebentrost2018quantum,Woerner2019,Egger2020} to pricing options\cite{Zoufal2019,martin2019pricing,kaneko2020quantum,PhysRevA.98.022321}. As mentioned previously, the most numerically tractable method of  option pricing is to use classical Monte-Carlo methods, which have  been extended to quantum settings in recent works. Ref\cite{Stamatopoulos_2020,martin2019pricing,PhysRevA.98.022321} pointed out that one can achieve a theoretical quadratic speedup $w.r.t$ number of Monte-Carlo samples with the use of Quantum Amplitude Estimation (QAE)\cite{brassard2002quantum} as a subroutine for measuring the expectation value of an evolved price. However, canonical QAE uses quantum phase estimation as a subroutine and is not practical for near term quantum processors due to  required circuit depth. Several methods using tools like maximum likelihood estimation   have been developed to make QAE and QMC more practical for NISQ era quantum devices\cite{Suzuki2020,brown2020quantum}.
For instance, Ref\cite{kaneko2020quantum} showed that one can, in practice, formulate a circuit that can be used to simulate the path dynamics of a given asset, but so far the proposed methods are not NISQ-friendly. Ref\cite{vazquez2020efficient} showed that one can create a NISQ efficient method to prepare the final probability state of the option price that obeys a Heston stochastic model. This method can in principle be extended to state preparation of any stochastic differential equation. In this paper, we propose an algorithm that  reformulates  the classical option pricing problem in the quantum setting. The key idea is to recast the pricing problem as a set of differential equations, which then maps to evolving a quantum wave function in imaginary time


We start in \cref{sec:fin} by briefly describing the option pricing problem and then reformulating it in the quantum setting. In \cref{sec:qalgo}, we outline the proposed algorithm which is made up of two independent parts which we discuss in \cref{sec:stateprep} and \cref{sec:imevolve}. We show numerical results to validate our proposal in \cref{sec:results} and finally end with a summary of our findings and suggestions for future work.

\section{Financial Hamiltonian} \label{sec:fin}
We start by considering a stock price $S(t)$, which is modeled as a random
stochastic variable obeying Geometric Brownian motion that satisfies the equation\cite{black1973pricing}
\begin{equation}
    dS(t)=\mu(t) S(t)dt+\sigma(t) S(t)d\mathcal{B}_t. \label{eq:1}
\end{equation}
Here $\mu(t)$ and $\sigma(t)$ are the expected return and volatility of the
stock which for the discussion, we will assume to be time independent
\textit{i.e.} $\mu(t) \approx \mu $ and  $\sigma(t)\approx\sigma$, and finally
$\mathcal{B}_t$ is the Brownian process satisfying
$\mathcal{B}_{t+s}-\mathcal{B}_{s}=\mathcal{N}(0,t)$ and $\mathcal{B}_0=0$. One
can then consider a function $O(S_t,t)$ dependent on a stochastic variable $S_t$ satisfying
\cref{eq:1} and time $t$, which satisfies the following time derivative due to Ito calculus,
\begin{equation}
    \frac{\mathrm{d} O}{\mathrm{~d} t}=\frac{\partial O}{\partial t}+\frac{1}{2} \sigma^{2} S^{2} \frac{\partial^{2} O}{\partial S^{2}}+\mu S \frac{\partial O}{\partial S}+\sigma S \frac{\partial O}{\partial S} \mathcal{B}\label{eq:2}.
\end{equation}
The Black-Scholes-Merton (BSM) model \cite{black1973pricing,merton1973theory} is then
obtained by removing the Brownian randomness of the stochastic process by
introducing a random process correlated to \cref{eq:2}. This is often termed
as \textit{hedging} and is done by constructing a a portfolio $\mathcal{P}$,
whose evolution is governed by
\begin{equation}
    \frac{d\mathcal{P}}{dt}=r\mathcal{P}\label{eq:3},
\end{equation}
where $r$ is the short term risk free interest rate. A possible choice of $\mathcal{P}$ is
$\mathcal{P}=O-S\left(\frac{\partial O}{\partial S}\right)$. In this context,
$O$ is called the \textit{option} and the portfolio is made up of an option and
an amount of underlying stock $S$ proportional to $\frac{\partial f}{\partial
S}$. Combining \cref{eq:2} and \cref{eq:3}, one finally arrives at the BSM
equation given by 
\begin{equation}
    \frac{\partial O}{\partial t}+\frac{1}{2} \sigma^{2} S^{2} \frac{\partial^{2} O}{\partial S^{2}}+r S \frac{\partial O}{\partial S}=r O \label{eq:4}.
\end{equation}

It is necessary to note that there are many assumptions underpinning the above result.
We have assumed a constant interest rate $r$, an option continuously re-balanced
within the portfolio, the absence of arbitrage and finally infinitesimal divisibility of
stocks with no transaction costs. Because of the first order and second order
nature of time, and asset value in  \cref{eq:4}  needs boundary
conditions \textit{w.r.t} $S$ and one boundary condition \textit{w.r.t} $t$. For
a European call option, the value of $O$ at expiry time $T$  is given by
$O(T,S)=max(S-K,0)$ $\forall S$ where $K$ is the exercise price. The second
condition involves the boundary that $O(t,S) \sim S$ as $S\rightarrow \infty$. And
finally, the boundary condition for time trivially involves $O(t,0)=0$ $\forall
t$. 

Without loss of generality, we now re-scale the two independent variables $(t,S)$
by $(\tau,x)$ defined by $t=T-\frac{2\tau}{\sigma^2}$ and $S=Ke^x$. To
explicitly see the analogy \textit{w.r.t} Quantum mechanics, we redefine
$O(t,S)=E\phi(\tau,x)$. This re-scaling changes \cref{eq:4} to
\begin{equation}
    \partial_t \phi = \partial_x^2 \phi + \left(\frac{2r}{\sigma^2} -1 \right)\partial_x \phi - \left(\frac{2r}{\sigma^2}\right) \phi \label{eq:5}
\end{equation}

Notice that one can now associate a quantum mechanical formalism and interpret
the option price $\phi:= \ket{\phi}$ in continuous basis $\ket{x}$, the underlying security price. This now elegantly extends the time dependence as $\phi(t,x)=\braket*{x}{\phi(t)}$ as the projection of the time evolved option price state. \cref{eq:5} thus reduces to the form $\partial_t \ket{\phi(t)} = \hat{H}_{BSM} \ket{\phi(0)}$, where the BSM Hamiltonian is given as $\hat{H}_{BSM}=\partial_x^2 +\left(\frac{2r}{\sigma^2} -1
\right)\partial_x  - \left(\frac{2r}{\sigma^2}\right) $. Note that the resulting Hamiltonian is not Hermitian because of the $\partial_x$ operator. It can be shown that this
Non-Hermiticity can be taken care of by making a similarity transformation\cite{bagarello2020susy} which finally results in an effective Hermitian Hamiltonian 
\begin{equation}
\hat{H}=e^{-\zeta x}\hat{H}_{BSM}e^{\zeta x}=-\frac{\sigma^2}{2}\partial_x^2 +\frac{(r+\sigma^2 /2)^2}{2\sigma^2} \label{eq:6},
\end{equation}
where $\zeta=\frac{1}{2} - \frac{r}{\sigma^2}$. Thus to find the option price at time $t$ which is given by the state $\ket*{\phi(t)}$, one starts by evolving the initial state $\ket*{\phi(0)}$ with the Hermitian Hamiltonian $\hat{H}$ in the Wick's rotated imaginary time space resulting in 
\begin{equation}
\ket*{\phi(t)}=\mathcal{A}(t)e^{t\hat{H}}\ket*{\phi(0)},\label{eq:7}
\end{equation}
where $\mathcal{A}(t)=(\bra*{\phi(0)}e^{2t\hat{H}}\ket*{\phi(0)})^{-1/2}$ is a normalization factor.

Although we have derived the European option pricing Hamiltonian, a similar Hamiltonian for Asian path-dependent options can be derived\cite{vecer2002unified} and is shown to be a time-dependent Hamiltonian given by 
\begin{equation}
\hat{H}_{Asian}=\frac{\sigma^2}{2}\bigl(\kappa(t) - x\bigr)^2\partial_x^2 \label{eq:8},
\end{equation}
where $\kappa=(rT)^{-1}(1-e^{-r(T-t)})$. It should be noted that both the Hamiltonian kernels resemble a heat equation and the deep analogy between stochastic equations and Heat/diffusion equations has been exploited in various contexts.

\section{Quantum Algorithm} \label{sec:qalgo}
\begin{figure}
    \includegraphics[width=\columnwidth]{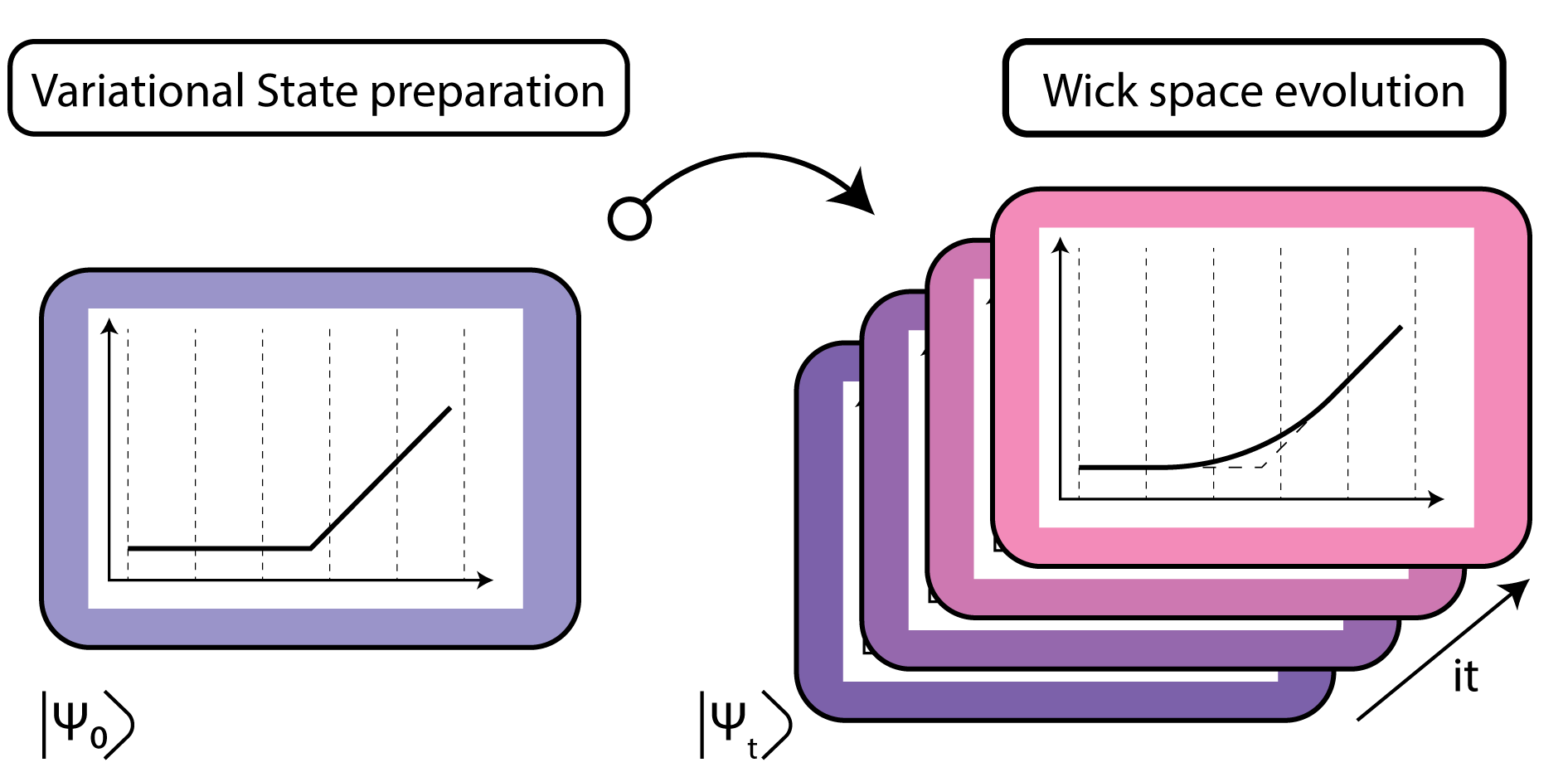}
    \caption{Proposed algorithm, where we start from initial distribution of the option price at $t=0$ with a variationally prepared state $\ket{\psi_0}$ and then evolve it in Wick rotated imaginary time space to find at any time $\ket{\psi_t}$} \label{fig:qalgo}
\end{figure}
We start with encoding the initial boundary conditions of the option price into initial quantum state, and then evolve it with the Hamiltonian corresponding to the pricing method in imaginary time. We define the basis of an n-qubit system $\in \mathbb{C}^{2^n}$ in independent orthogonal states $\ket*{x}$ spanning the space of underlying security prices discretized into $2^n$ pieces. We then write $\ket*{\phi(x,t)}$ ,the value of the option price when the underlying asset is at price $x$ at time $t$ until maturity, as $\ket*{\phi(x,t)}= \sum_x \sqrt{p_x(t)} \ket*{x}$, where $\sum_x p_x(t)=1$ for all time $t$, after which the results are finally scaled to get the actual cost. We start with the boundary condition that $p_x(0)=\alpha_0\text{max}(e^x-K,0) e^{(\frac{r}{\sigma^2} - \frac{1}{2})x}$, where $\alpha_0$ is the normalization factor. As mentioned before, for a system with Hamiltonian $H$, evolving in real time  $t$, the time propagator is given by $e^{-iHt}$. The corresponding propagator in Wick rotated imaginary time ($\tau=it$) is given by the propagator $e^{-\tau H}$ which is  a non-unitary operator. Therefore, our algorithm consists of two major parts, preparing the initial state and Wick rotated imaginary time  evolution.

\section{Variational State preparation}\label{sec:stateprep}

\begin{figure}[h]
    \includegraphics[width=\columnwidth]{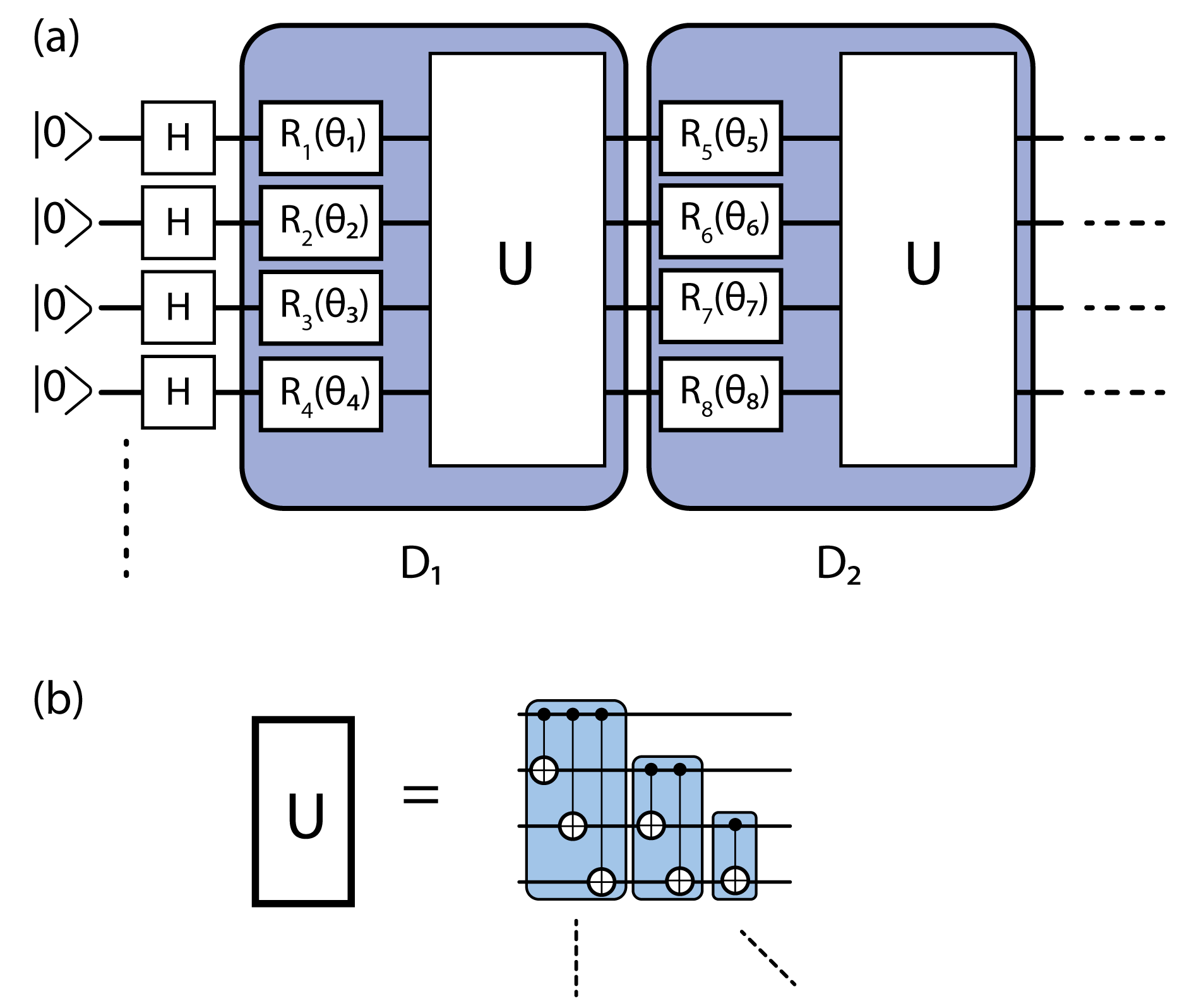}
    \caption{(a) Circuit anzats for Variational Hilbert space evolution with each depth having randomly rotated chosen gates as described in the text and two qubit entanglement gates $U$ defined in (b)}\label{fig:variational_anzats}

\end{figure}
 The first step of the algorithm involves the preparation of the initial state $\ket{\psi_0}$  shown in \cref{fig:qalgo}. We start by using a parameterized quantum circuit to varaitionally prepare the initial state of the system. Expressivity of such a parameterized quantum circuit depends on the number of parameters used; more parameters generally lead to higher variational flexibility, but come with the cost of having not-so-NISQ friendly circuits (\textit{i.e} deeper quantum circuits) and are often plagued by difficultly in optimization \cite{McClean_2018}. The goal of finding parameterized circuits which minimize the circuit depth and maximize the accuracy and expressivity has led to a number of approaches such as hardware efficient ansatz, physically motivated fixed ansatz, adaptive ansatz \textit{etc.}. 
 
 Although any circuit that is expressive enough to represent the initial boundary condition $\ket{\psi_0}$ can be used, we choose two specific PQCs that are suitable for each subroutine mentioned in the subsequent sections. For the Variational Hilbert space evolution, we start with superimposing the state by applying Hadamard gate, after which we make $d$ layer unitary $D_j$ ($j\in \{1,2\ldots d\}$) given by sets of single qubits and randomly chosen rotation gates $R_\alpha (\theta_i)$ where $\alpha \in \{X,Y,Z\}$ and $i \in \{1,2,3\ldots N\}$, followed by a two qubit entanglement layer $U$ consisting of pairwise $CNOT$s. The resulting circuit has $nd$ parameters that are to be optimized and is shown in \cref{fig:variational_anzats}, where $n$ is the number of qubits.
 \begin{figure}[h]
    \includegraphics[width=6cm]{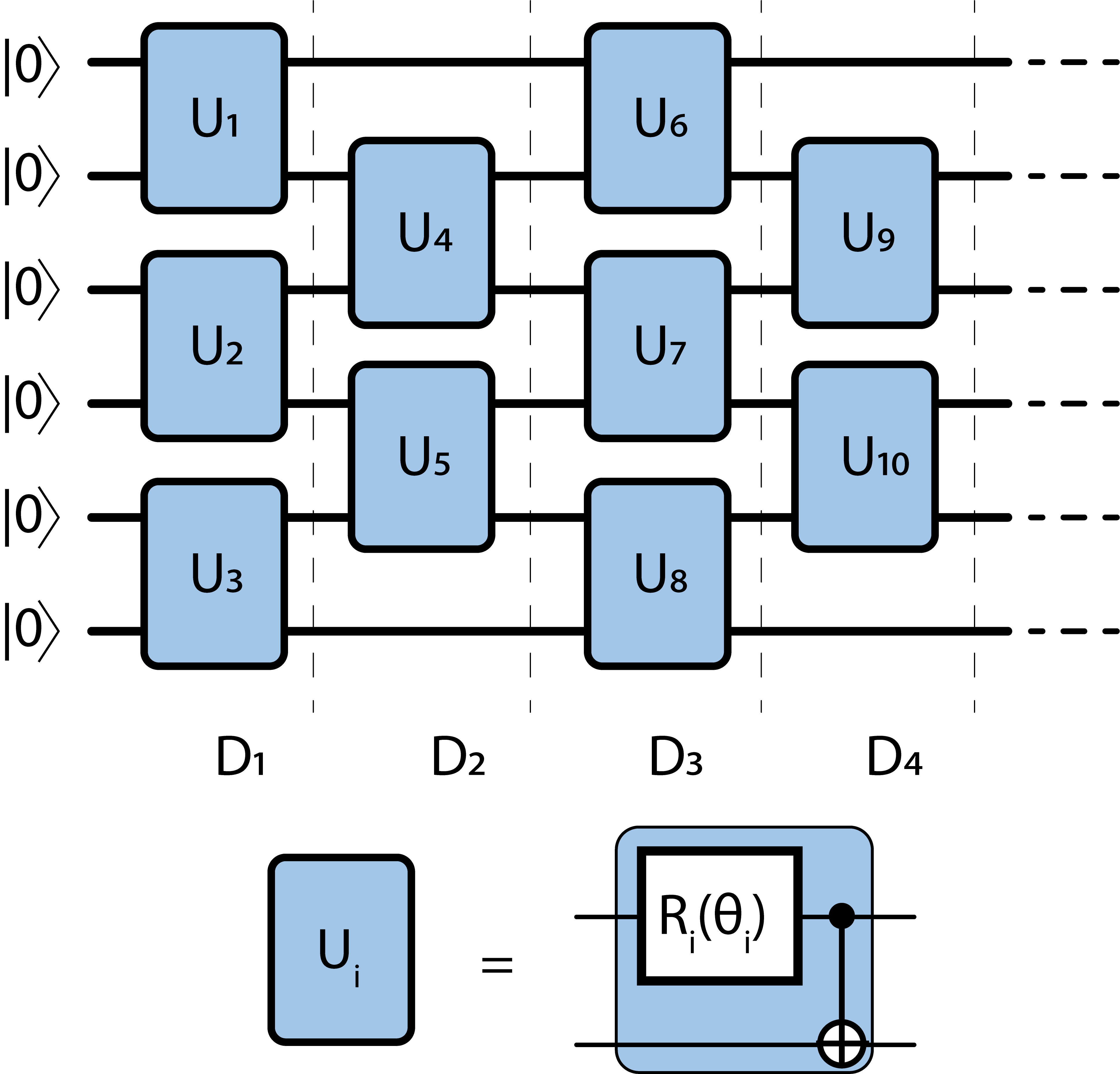}
    \caption{Paramatric quantum circuit (PQC) for hardware efficient variational anzatz. }\label{fig:pqc}
\end{figure}

 For the second hardware efficient method, we chose an ansatz as shown in \cref{fig:pqc}. Each depth unitary at depth $i$ ($D_i$) is made up of the nearest neighbor two qubit entanglement gates ($U_j$) starting from either the first or second qubit. $U_j$ is made up of randomly chosen single qubit rotation gate $R_\alpha (\theta_j)$ where $\alpha \in \{X,Y,Z\}$ and $j \in \{1,2,3\ldots N\}$ followed by a $CNOT$. This is more hardware efficient than the previously introduced ansatz since only nearest  neighbour qubit connectivity is required as opposed to an arbitrary qubit connectivity, which in turn reduces the total amount of $CNOT$ gates and circuit depth. 

 Imaginary time evolution is often used as a tool for driving an initial wave function towards the ground state of a given system. In that case, as long as errors due to an imperfect ansatz do not cause the evolution to become trapped in local minima, one does not care about the path of true imaginary time evolution, as ultimately, the system will always be driven towards the ground state. By contrast, our anzatz does not only need to be expressive enough to represent the initial and final state, but also the intermediate states, thus building anzats circuits that are well suited to imaginary time evolution is an intersting problem that will be addressed in future work.


\section{Wick space evolution}\label{sec:imevolve}
In this section we describe two algorithms that can be used to imaginary time evolve the financial Hamiltonians. The first method, detailed in \cref{subsec:vhse} involves approximating the state vector using a variational parametrized ansatz which is then evolved by evolving the parameters of the ansatz circuit. In the second method, we use variational algorithm similar to the Variational Quantum Eigensolver introduced in Ref\cite{benedetti2020hardwareefficient}, where a hardware efficient variational circuit is used to simulate individual terms in the Hamiltonian by minimizing a cost function. This is explained in \cref{subsec:hevs}. 
\subsubsection{Variational Hilbert space evolution} \label{subsec:vhse}
\begin{figure}
    \includegraphics[width=\columnwidth]{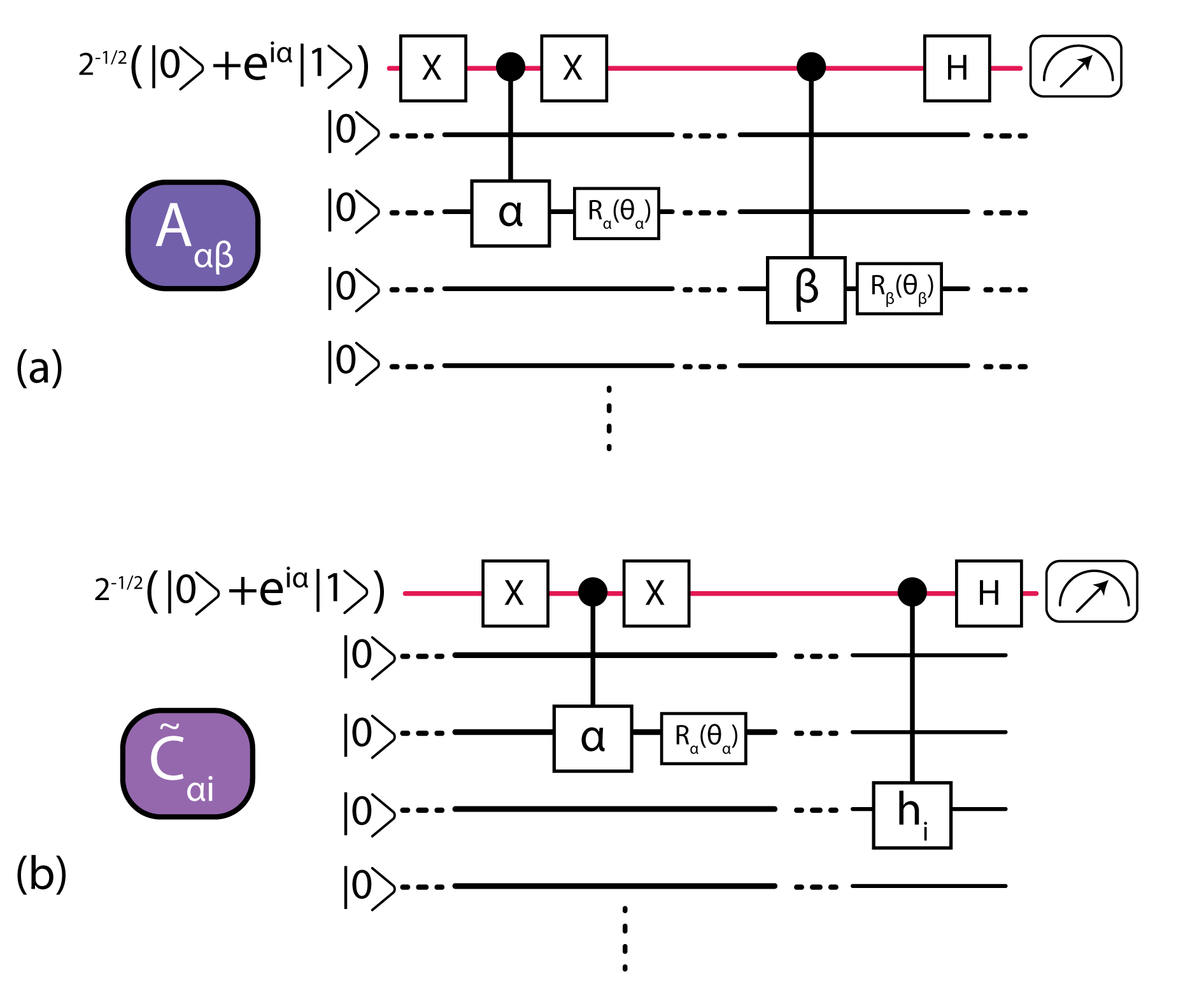}
    \caption{  Circuit for calculating (a) $A_{\alpha \beta}$ (b) $\tilde{C}_{\alpha i}$} up-to a factor. The red qubit is the ancilla qubit where the final measurement is made.\label{fig:aij_cij}
\end{figure}
Following \cite{endo2020variational}, we started by first writing out finance Hamiltonians as $\hat{H}=\sum_i \lambda_i h_i$ with real coefficients $\lambda_i$ and observables $h_i$, that are tensor products of Pauli matrices. Our Hamiltonian contains operators of the form $\partial_x^2$ , which are then converted using finite difference methods, converted to a tri-diagonal form in the $x$ subspace while operators of the form  $f(x)$ are trivially diagonal. This is shown in detail in \cref{sec:app:ham}. In the variational Hilbert space method, instead of directly calculating the state $\ket*{\phi(t)}$, we approximate it using an $N-$dimensional variational circuit made up of parameters $\vect{\theta}(t)=\{\theta_1,\theta_2\ldots\theta_N\}$ giving us $\ket*{\phi(t)}\approx\ket*{\psi(\vect{\theta}_N(t))}$, using a variational quantum anzats as shown in \cref{fig:variational_anzats} and discussed in \cref{sec:stateprep}. Thus our entire circuit can be written as $\ket*{\psi(\vect{\theta}(0))}=V(\vect{\theta}(0))\ket*{0}$. We start by variationally finding $\vect{\theta}(0)$ that best approximates our initial condition $\ket*{\phi(0)}$. We then use MacLachlan's variational principle\cite{mclachlan1964variational} to translate the time propagation of state $\ket*{\phi(t)}$ to time evolve the variational vector $\vect{\theta}(t)$, which in turn propagates $\ket*{\psi} $. MacLachlan's variational principle is given by 
\begin{equation}
\delta \|\left(\partial / \partial t+H-E_{\tau}\right)|\phi(t)\rangle \|=0 \label{eq:9}
\end{equation}
 where $\||\psi\rangle \|=\sqrt{\langle\psi \mid \psi\rangle}$ is the norm of the quantum state $\ket*{\psi}$. Following the same for $\ket*{\psi(\vect{\theta})} $, one gets the first order equation of $\vect{\theta}(t)$ that dictates the evolution of the parameters, which is given by \cite{yuan2019theory}

\begin{equation}
 \sum_j A_{ij}\frac{d\vect{\theta}_j}{dt}=C_i \label{eq:10},
 \end{equation}
where:
\begin{align}
 A_{ij} &= \Re\left(\frac{\partial\langle\psi(t)|}{\partial \theta_{i}} \frac{\partial|\psi(t)\rangle}{\partial \theta_{j}}\right)\label{eq:11}  \\
 C_i &= \Re\left(-\sum_{\alpha} \lambda_{\alpha} \frac{\partial\langle\psi(t)|}{\partial \theta_{i}} h_{\alpha}|\psi(t)\rangle\right)\label{eq:12},
\end{align}

Here $h_\alpha$ and $\lambda_\alpha$ are the Hamiltonian terms. The idea behind the variational Hilbert space evolution algorithm comes from the fact that quantities $A_{ij}$ and $C_i$ can be efficiently measured using a quantum circuit as shown in Ref\cite{li2017efficient}. This is because $\frac{\partial|\psi(t)\rangle}{\partial \theta_{i}}=f_{i} \tilde{V}_{i}|\overline{0}\rangle$ where $\tilde{V}_{\alpha}=R_{N}\bigl(\theta_{N}\bigr) \ldots R_{\alpha+1}\bigl(\theta_{\alpha+1}\bigr) R_{\alpha}\bigl(\theta_{\alpha}\bigr) \pmb{\hat{\alpha}} \ldots R_{2}\bigl(\theta_{2}\bigr) R_{1}\bigl(\theta_{1}\bigr)$ with $\pmb{\hat{\alpha}}$, the Pauli gate corresponding to $R_{\alpha}\bigl(\theta_{\alpha}\bigr)$ and $f_\alpha=-\frac{i}{2}$ where $i=\sqrt{-1}$. Hence, the calculation of \cref{eq:11} and \cref{eq:12} reduces to 

\begin{align} 
A_{\alpha \beta} &=\Re\left(f_{\alpha}^{*} f_{\beta}\left\langle\overline{0}\left|\tilde{V}_{\alpha}^{\dagger} \tilde{V}_{\beta}\right| \overline{0}\right\rangle\right) \label{eq:13},\\
C_{\alpha} &=\Re\left(\sum_{i} f_{\alpha}^{*} \lambda_{i}\left\langle\overline{0}\left|\tilde{V}_{\alpha}^{\dagger} h_{i} V\right| \overline{0}\right\rangle\right) \label{eq:14},\\
&=\sum_i \tilde{C}_{\alpha i}.
\end{align}

Each of the above terms $A_{\alpha \beta}$ can be calculated using the circuit shown in \cref{fig:aij_cij}(a) while $\tilde{C}_{\alpha i}$ terms are calculated by circuits shown in \cref{fig:aij_cij}(b) (up-to a factor). Note that both the circuits have an extra ancilla which essentially measures the real part of each term. Here $\alpha$ is  $\arg(f_{\alpha}^{*} f_{\beta})$ for $A_{\alpha \beta}$ and $\arg(f_{\alpha}^{*} \lambda_{i})$ for $\tilde{C}_{\alpha i}$.

Having calculated $A(t)$ and $\vect{C}(t)$ at time $t$, one can invoke \cref{eq:10} to calculate the updated $\vect{\theta}(t)$ as $\vect{\theta}(t+\delta t)\approx \vect{\theta}(t)+A^{-1}(t)\dotproduct\vect{C}(t)\delta t$. We have used the Euler finite difference formula for approximating \cref{eq:10}, where more sophisticated numerical approximations and higher order terms as discussed in \cref{sec:app:ham} can very well be used to increase accuracy. Hence by propagating $\vect{\theta}(t)$ in time, we generate $\ket{\psi(t)}$. Once we have a set of time depedent $\vect{\theta}(t)$s, we can plug them back to the initial parameterized quantum circuit to extract the option price at these discretized times.

\subsubsection{Hardware efficient variational evolution}\label{subsec:hevs}
By using the ideas introduced in Ref\cite{benedetti2020hardwareefficient}, we want to propagate our hamiltonian $\hat{H}=\sum^K_i \lambda_i h_i$. Note that $h_i$'s, in general, need not commute with each other. We then approximate the evolution of state $|\psi(0)\rangle$ by expanding $e^{-i\hat{H} \tau}$ as a Trotter product
\begin{widetext}
\begin{equation}
    \begin{aligned} e^{-i \tau \hat{H}}|\psi\rangle & \approx\left(e^{-i \zeta \lambda_{K} h_{K}} \ldots e^{-i \zeta \lambda_{1} h_{1}}\right)^{N}|\psi\rangle\\ &=e^{-i \zeta \lambda_{K} h_{K, N}} \ldots e^{-i \zeta \lambda_{1} h_{1, N}} \ldots e^{-i \zeta \lambda_{K} h_{K, 1}} \ldots e^{-i \zeta \lambda_{1} h_{1,1}}|\psi\rangle, \end{aligned}\label{eq:16}
\end{equation}
\end{widetext}

with $\tau=it$ for imaginary time evolution and $\zeta=\frac{\tau}{N}$ for dividing the time into $N$ steps. $h_{i,j}$ refers to the $h_i^{th}$ term applied in $j^{th}$ time step. The first step involves variationally approximating the initial state $|\psi(0)\rangle$, which is given by $\ket{\psi(0)}\approx\ket{\psi(\vect{\theta}_0)}$, then the first application of $e^{-i \zeta \lambda_{1} h_{1,1}}$ on state $\ket{\psi(\vect{\theta}_0)}$ results in $\ket{\psi(\vect{\theta}_1)}$ with $\vect{\theta}_1=\underset{\vect{\alpha}}{\mathrm{argmin}}$ $C_0(\vect{\alpha})$ where 
\begin{equation}
    C_0(\vect{\alpha})=\Bigl|\ket{\psi(\vect{\alpha})}-e^{-i \zeta \lambda_{1} h_{1,1}}\ket{\psi(\vect{\theta}_0)}\Bigr|\label{eq:17}.
\end{equation}
It can be shown that\cite{benedetti2020hardwareefficient} minimization of $C_0(\vect{\alpha})$ is the same as maximization of $\operatorname{Re}\bigl(\bra{\psi(\vect{\theta}_0)}e^{-i \zeta \lambda_{1} h_{1,1}}\ket{\psi(\vect{\alpha})}\bigr)$

Thus generalizing the above discussion, for the $n^{th}$ term in \cref*{eq:16}, the state vector after applying $n$ terms in the Trotter product is variationally obtained by calculating the angles that maximize $\mathcal{F}_n\left(\vect{\alpha}\right)$ given by 
\begin{equation}
    \mathcal{F}_{n}\left(\vect{\alpha}\right)=\operatorname{Re}\Bigl(\left\langle\psi(\vect{\theta}_{n-1})\left|e^{i \zeta \lambda_{n} h_{n}}\right| \psi\left(\vect{\alpha}\right)\right\rangle\Bigr)\label{eq:18}
\end{equation}

where $\vect{\theta}_{n-1}$ is the optimal angle in the previous $(n-1)^{th}$ step. Further, using the fact that $e^{-t\lambda_ih_i}=\cosh(t \lambda_i)\mathbf{1}-\sinh(t \lambda_i)h_l$, we can further reduce our maximization function \cref*{eq:18} as
\begin{multline}
    \mathcal{F}_{n}\left(\vect{\alpha}\right)=\cosh(t \lambda_i) \operatorname{Re}\Bigl(\left\langle\psi(\vect{\theta}_{n-1})| \psi\left(\vect{\alpha}\right)\right\rangle\Bigr) \\   -\sinh(t \lambda_i)\operatorname{Re}\Bigl(\left\langle\psi(\vect{\theta}_{n-1})\left|e^{i \zeta \lambda_{n} h_{n}}\right| \psi\left(\vect{\alpha}\right)\right\rangle\Bigr)\label{eq:19}
\end{multline}

It was shown in Ref\cite{PhysRevResearch.2.043158,vidal2018calculus,parrish2019jacobi,ostaszewski2019quantum} that parameterized circuits of the type shown in \autoref{fig:variational_anzats} have a very efficient coordinate-wise optimization algorithm that eliminates the computation of costly gradients. Essentially, one finds the optimal angle for one gate while fixing all others to their current values, and sequentially cycles through all gates. This is easy to carry out, as in each step, the energy has a sinusoidal form with period $2\pi$. For the $k^{th}$ gate in the circuit, one obtains the update rule for finding the angle  $\alpha_k^*$ as 
\begin{equation}
    \begin{aligned} \alpha_k^{*} &=\arg \max _{x} f_{n, k}(x) \\ &=\frac{\pi}{2}-\arctan \left(\frac{f_{n, k}\left(\alpha_k\right)}{f_{n, k}\left(\alpha_k+\frac{\pi}{2}\right)}\right)+\alpha_k \end{aligned}
\end{equation}
where $f_{n, k}(x)=\mathcal{F}_{n}\left(\dots \alpha_{k-1},x,\alpha_{k+1}\dots\right)$ with the current parameter $\alpha_k$.

As noted by Ref\cite{benedetti2020hardwareefficient}, the calculation of terms in \cref{eq:19} can easily be carried out out in a Hardware efficient manner reducing both the number of CNOTs and ancilla qubits as compared to the  method outlined in  \autoref{subsec:vhse}. This new method also avoids the need to calculate matrix inversion.

\section{Results}\label{sec:results}
\begin{figure}
    \includegraphics[width=\columnwidth]{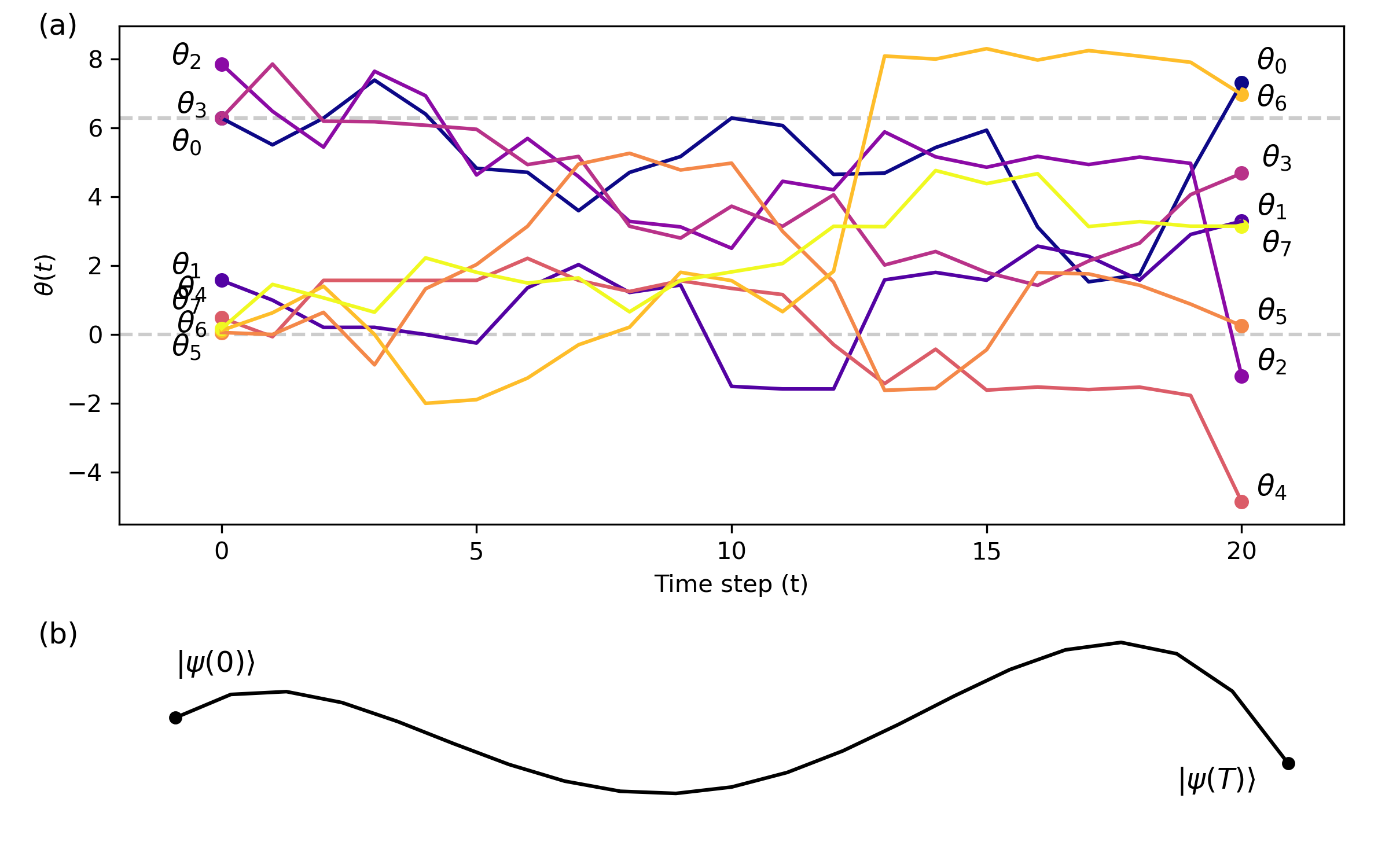}
    \caption{(a) Time evolution of parameters given by update rule (b) Illustration of corresponding time evolution of the state vector  }\label{fig:angles}
\end{figure}

In this section, we show the numerical results obtained using the Variational Hilbert space evolution algorithm described in \cref{subsec:vhse}. For simplicity, we calculate the option price using a varational Hilbert space evolution subroutine, although this can be replaced by the hardware efficient evolution. To illustrate the evolution of the wave function by evolving the variational parameters, we use a 4 qubit system. Restricting ansatz depth to $2$, we have $8$ variational parameters that are time evolved using \cref{eq:10}. This is shown in \cref{fig:angles}(a), where we plot the time evolution of each parameter for 20 Trotter steps. Even though our goal is to evolve the quantum state by a small time step in its Hilbert space (shown as illustration in \cref{fig:angles}(b)), one can see from \cref{fig:angles}(a) that the change in variational parameters for each individual time step is neither small nor trivial. This is associated with the deep connection to quantum generalization of \textit{natural gradients} as explained in Ref.\cite{Stokes_2020}. \cref{eq:10} essentially relates the non-trivial relationship between a translation step in parameter space, and the corresponding translation in the problem space, which in our case is the Hilbert space. The 2D tensor $A$ in the update rule is the well known Fubini-Study metric tensor \cite{wilczek1989geometric,hackl2020geometry,koczor2019quantum} which accounts for the non-uniform  effect of the parameter's gradients on the quantum states. 

\begin{figure}
    \includegraphics[width=\columnwidth]{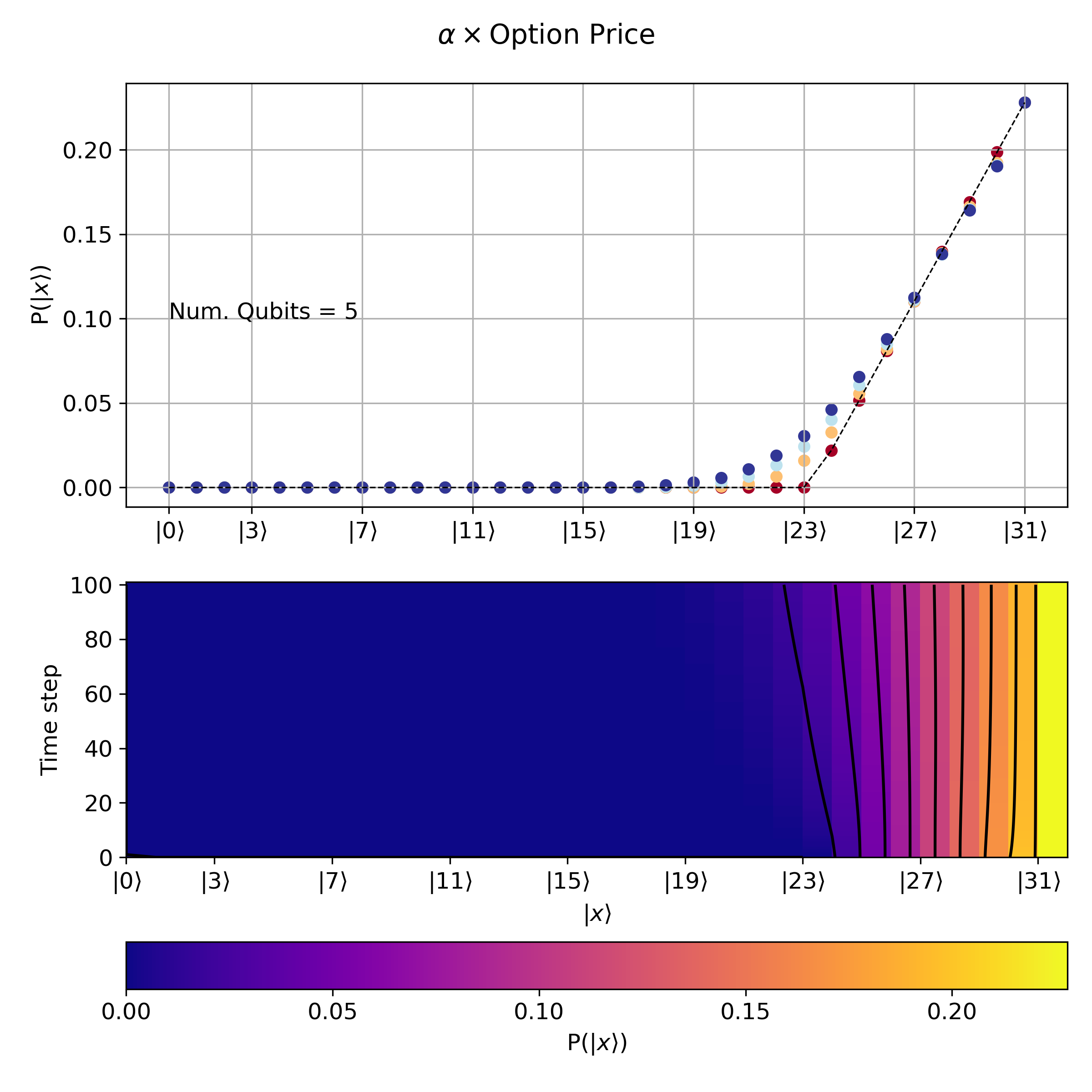}
    \caption{(a) Evolution of the option price from time $t=0$ (red) to $t=1$ (blue). (b) Same as (a) with  colorplot showing the change in option price at around the wave vector $\ket{23}$}\label{fig:option_main}
\end{figure}

Moving beyond the evolution of parameters to better simulate the option pricing, we now use a 5 qubit system with depth $4$ ansatz to time evolve the system. The result of this simulation is shown in \cref{fig:option_main}. We decompose our final time step $T=1$ into $100$ Trotter steps and thus our time and spatial discretization is given by $N_T\times N_X=100\times 2^5$. In the top part of the figure, we plot the probability of each state at different time steps indicated by the color of the plot with red being $t=0$ and blue being $t=T=1$, and where the dotted line shows the initial starting state. The actual option value corresponding to each price is then obtained by scaling the probability to arbitrary units. The lower part of the figure shows the same data plotted in 2D with the price on the $x$ axis, the time step on the $y$ axis, and probability/scaled option value shown as a color map. One can now look more closely at the non linear effect of time evolution on the option pricing around state $\ket{23}$.


Here we want to discuss possible sources of errors in the algorithm. There are five main types of errors that propagate within the algorithm. 1. Errors due to limited expressivity of the variational circuit in describing the trial wave function. 2. Errors that come from approximating the Hamiltonian and Troterization. 3. Errors in the approximate numerical integration methods employed in \cref{eq:10}, which is always going to be approximate because of the finite discretization of both time and space. 4.Shot noise in measuring equation coefficients $A$ and $V$ and finally 5.Errors due to noise in the quantum machines. Errors due to variational anzats can be systematically improved by making the circuit more expressive by increasing the depth and also by adapting better techniques to optimize the initial parameters. Errors due to Troterization can be reduced by adding higher order trotter terms. While we have used a simple euclidean finite difference method for integrating \cref{eq:10}, more sophisticated techniques like Adams-Moulton or Newton-Cotes can be used to drastically reduce errors. 

\begin{figure}
    \includegraphics[width=\columnwidth]{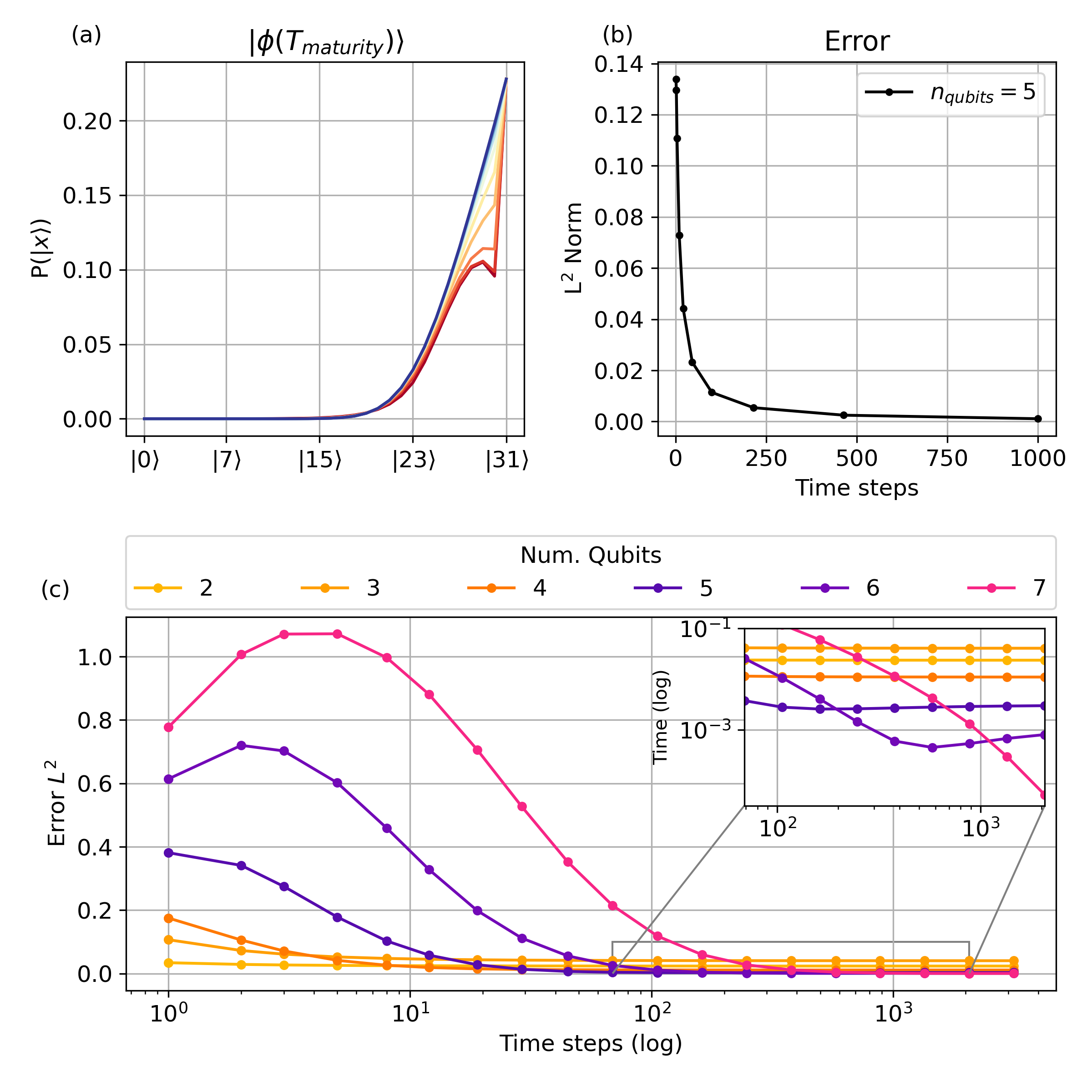}
    \caption{(a) Option price at maturity time $T=1$ as function of number of time steps with red $=1$ time step and blue $=10^3$ time steps for 5 qubits (b) Error denoted by $L^2$ norm $w.r.t$ analytical solution for 5 qubits. (c) Error as calculated from $L^2$ norm for various qubits/spatial discretization and $log$ time discretization. Inset shows the zoomed in portion in $log-log$ plot }\label{fig:option_error}

\end{figure}

Errors due to finite time step descritization are found to be sensitive and closely coupled to the space descritization, \textit{i.e} number of qubits. This is to be expected as the entire algorithm boils down to solving a finite difference solution of the classical \cref{eq:10}. Classical finite difference solutions have errors given by$ \mathcal{O}(N_T^{-1}) + \mathcal{O}(N_X^{-2})$, where $N_T$ is the number of time steps and $N_X$ is the number of spatial bins. This is because of the first order and second order nature of the differential equation involved. Thus the time step must be quadratically smaller than the spatial step for comparable accuracy. This is demonstrated in \cref{fig:option_error}, where in (a) for a given space descritization of $2^5$, we plot the option price at maturity time $T=1$ for various time descritization steps with red being $N_T=1$ and blue being $N_T=10^3$ (each curve independently normalized). In (b) we plot the error as measured by $L^2$  norm distance from the analytic solution as a function of $N_T$. As discussed earlier, accuracy of the final result can be systematically increased by either increasing $N_T$ or $N_X$, but in a non-trivial manner as discussed in \cref{sec:app:ham}. We show this in (c), where we plot the mean squared error instead of $L^2$ for various $N_T$ and numbers of qubits (or equivalently $N_X$). This shows the non-trivial relationship between the errors as for a given $N_T$ one needs to have a sufficiently large $N_X$ to see systematic improvement in error as one increases $N_X$. For instance, close to $N_T\approx 10$, one sees that going from 2 qubits to 7 qubits increases the errors, while there is a threshold around $N_T\approx 10^2$ after which the error of a 7 qubit space discretization is smaller than that of a 2 qubit one (shown in the inset). Nonetheless, as $N_T\rightarrow \infty$ , $\epsilon(N_{X_1},N_{N_T\rightarrow \infty})<\epsilon(N_{X_2},N_{N_T\rightarrow \infty})$ for $N_{X_1}<N_{X_2}$ where $\epsilon(N_X,N_T)$ is the error due to spatial and time discretization. This can also be seen by comparing the final values of the error shown in the inset, where the errors are arranged such that the system with the largest number of spatial steps has the least error, as expected. 



\section{Conclusion}

In this work, we have shown  that one computationally resource intensive but industrially relevant class of financial problems, \textit{i.e} option pricing, can be reformulated as a quantum imaginary time evolution of a wave function problem. Quantum  Monte Carlo methods offer a theoretical speed up to this problem, but are still impractical for near term quantum devices due required unattainable circuit depth.  The importance of our work lies in transferring the solution of this computationally resource intensive financial problem into a hybrid classical/quantum domain where it may become practical for near term quantum devices. We have presented  a proof of concept demonstration for European option pricing problem, but it can be extended to other option pricing problems such Asian options. 


The proposed algorithm consists of two main parts, namely the initial state preparation and the imaginary time propagation. For the imaginary time propagation we described two different algorithms, namely variational Hilbert state evolution and the Hardware efficient variational evolution. 
After outlining the basic mechanisms of both of the imaginary time evolution algorithms, we selected the former  method for numerically evaluating the value of the option price as proof of concept. While the method used for numerical simulation (\textit{Variational Hilbert space evolution} \cref{subsec:vhse}) can be considered NISQ friendly, it still requires   arbitrarily distant qubit connectivity  and  numerical matrix inversion with dimensions that scale \textit{w.r.t} required accuracy/free parameters. This is overcome by the  \textit{Hardware efficient variational evolution} algorithm (\cref{subsec:hevs}), which includes only the nearest neighbor  connectivity and  does not require matrix inversion. Presented algorithms   can systematically be tuned for accuracy; either by increasing the number of qubits or by increasing the number of parameters without dramatic change in depth of the circuits. 

A good choice of variational ansatz is crucial for the proposed method. Future work could focus on further exploration of better anzats for parameterized quantum circuits for state preparations that are specifically tuned and inspired for option curves. For example, one could find shorter depth circuits that might not be expressive in exploring the entire Hilbert space, but do occupy just the space needed to represent option prices for the time span of interest. Other imaginary time evolution methods, such as the one introduced in Ref\cite{Motta2019,gomes2020efficient,yao2020adaptive} could be explored further as an alternative to the method presented above.    

\section{Acknowledgement}

We thank Oktay Goktas, Edwin Tham, Faiyaz Hasan, Cara Alexander and Jalani Kanem for useful discussions regarding both quantum and classical option pricing. We are grateful to Elliot Macgowan for his useful feedback on the manuscript. Numerical results in this study were performed using IBM Quantum’s Qiskit SDK\cite{Qiskit}. We were informed of similar work done in Ref\cite{kubo2020variational}, while preparing the manuscript.

\appendix*
\section{Hamiltonian approximation}\label{sec:app:ham}

As shown in \cref{eq:6} and \cref{eq:8} evaluation of $\hat{H}_{BSM}$ and $\hat{H}_{Asian}$ involves evolution of the kinetic energy operator $\partial^2$. Given a small number $\epsilon>0$, the derivative of order $m$ for a singlevariate function $f$ satisfies the equation\cite{hildebrand1987introduction}
\begin{dmath}
    f^{(m)}(x) = \sum_{i=i_{\min }}^{i_{\max }} \frac{N_{i} f(x+i \epsilon)}{D_{i} \epsilon^{m}}+O\left(\epsilon^{p}\right)\label{eq:a1},
\end{dmath}

where $p>0$,  $i_{min}<i_{max}$ and  $N_i$, $D_i$ are integers with greatest common divisor $=1$. This leads to three main categories of numerical approximations. Forward-difference methods, where  $i_{min}>0$. Backward-difference methods, when $i_{max}<0$ and finally mixed difference when $i_{min}<0<i_{max}$. Note that in the case of forward difference and back difference, the resulting approximate differential matrix is bound to be not symmetric and hence non hermitian. Although non hermitian Hamiltonians can be simulated by variational hilbert space evolution algorithm\cite{endo2020variational}, we here choose the simpler mixed difference approximation where the resulting differential operator can be made hermitian. We choose the special choice that $i_{max}=-i_{min}=1$, which is often referred to as central difference method. We use $\partial_x f(x)\approx \frac{1}{\epsilon}\bigl(f(x+\epsilon)-f(x)\bigr) +\mathcal{O}(\epsilon)$, to reduce the discretized form of kinetic energy operator
\begin{dmath}
    \frac{\partial^2}{\partial x^2}f(t_i,x_j)\approx\frac{f(t_i,x_{j+1})-2f(t_i,x_j)+f(t_i,x_{j-1})}{\Delta x^2}+\mathcal{O}(\Delta x^2)\label{eq:a2},
\end{dmath}
where $\epsilon=\Delta x=\abs{x_{j+1}-x_j}$. This results in our Hamiltonian being in hermitian tridiagonal form given by

\begin{equation}
    \approx\epsilon^{-2}\begin{pmatrix}
        \frac{\alpha}{\epsilon^{-2}} & 0 &0& \dots & 0 \\
        1&-2&1 &\dots & 0 \\
        0&1&-2&1 &\vdots  \\
        \vdots & \vdots & \vdots & \ddots&\vdots \\
        0 & 0 & \dots & 0&\frac{\alpha}{\epsilon^{-2}} 
      \end{pmatrix}+\mathcal{O}(\epsilon^2)\label{eq:a3},
\end{equation}
where $\alpha$ is the required boundary condition. Note that one can systematically improve this approximation by increasing $i_{max}$ and $i_{min}$.  This comes at the cost of increasing the number of terms in the pauli decomposition and thus increases the size of the matrix $A$ and $C$ in \cref{eq:10}. For example, one can make $\mathcal{O}(\epsilon^4)$ approximation by setting $i_{max}=i_{min}=2$ which results in pentadiognal band matrix $(\hat{D})$ with diagonal elements $\hat{D}_{i,i}=-30$ and off diagonal elements given by $\hat{D}_{i,i-2}=\hat{D}_{i,i+2}=-1$ and $\hat{D}_{i,i-1}=\hat{D}_{i,i+1}=16$. Note that the form of \cref{eq:a1} forces all resulting approximations to result in sparse matrices which are computationally tractable. Further more, the same univariate derivative euler approximation is also used in approximating time derivative\cref{eq:10} resulting in errors of oder $\mathcal{O}(\Delta t)$. Thus in the classical case, one need to choose time and spatial discretization that approximately satisfies $\Delta t \approx \Delta x^2$ to have similar orders of accuracy.
\bibliography{finance,qc} 

\end{document}